\documentclass[fleqn,usenatbib]{mnras}

\usepackage{newtxtext,newtxmath}

\usepackage[T1]{fontenc}
\usepackage{ae,aecompl}

\usepackage{graphicx}	
\usepackage{amsmath}	

\usepackage{hyperref}
\usepackage{amssymb}	

\newcommand{\kms}{km s$^{-1}$}
\newcommand{\MgII}{\mbox{Mg\,{\sc ii}}}
\newcommand{\redm}{redMaPPer}
\newcommand{\redl}{$\lambda_{\rm red}$}
\newcommand{\redz}{$z_{\rm red}$}
\newcommand{\reds}{$\sigma_{\rm red}$}
\newcommand{\zqso}{$z_{\rm \scriptscriptstyle QSO}$}
\newcommand{\zabs}{$z_{\rm abs}$}
\newcommand{\ewa}{$EW_{\lambda2796}$}
\newcommand{\ewb}{$EW_{\lambda2803}$}
\newcommand{\ratio}{$EW_{\lambda2796}/EW_{\lambda2803}$}
\newcommand{\disps}{$\sigma_{\small \mbox{Mg\,{\sc ii}}}$}

\title[\MgII\ absorbers and galaxy clusters]{
Searching for \MgII\ absorbers in and around galaxy clusters}

\author[J. C. Lee et al.]{
Jong Chul Lee$^{1}$\thanks{E-mail: jclee@kasi.re.kr},
Ho Seong Hwang$^{1,2}$\thanks{E-mail: hhwang@astro.snu.ac.kr} 
and Hyunmi Song$^{3}$\thanks{E-mail: hmsong@yonsei.ac.kr} 
\\
$^{1}$Korea Astronomy and Space Science Institute, 
776 Daedeokdae-ro,  Yuseong-gu,  Daejeon 34055,  Korea\\
$^{2}$Astronomy Program,  Department of Physics and Astronomy, 
Seoul National University,  1 Gwanak-ro,  Gwanak-gu,  Seoul 08826,  Korea\\
$^{3}$Department of Astronomy,  Yonsei University, 
50 Yonsei-ro,  Seodaemun-gu,  Seoul 03722,  Korea\\
}

\date{Accepted XXX. Received YYY; in original form ZZZ}

\pubyear{2020}

\begin{document}
\label{firstpage}
\pagerange{\pageref{firstpage}--\pageref{lastpage}}
\maketitle

\begin{abstract}
To study environmental effects on the circumgalactic medium (CGM),
 we use the samples of \redm\ galaxy clusters, 
 background quasars and cluster galaxies from the SDSS.
With $\sim$82 000 quasar spectra,
 we detect 197 \MgII\ absorbers 
 in and around the clusters.
The detection rate per quasar is 2.7$\pm$0.7 times higher 
 inside the clusters than outside the clusters,
 indicating that \MgII\ absorbers are relatively abundant in clusters.
However, when considering the galaxy number density,
 the absorber-to-galaxy ratio is rather low inside the clusters.
If we assume that \MgII\ absorbers are mainly contributed 
 by the CGM of massive star-forming galaxies,
 a typical halo size of cluster galaxies is smaller than that of field galaxies 
 by 30$\pm$10 per cent.
This finding supports that
 galaxy haloes can be truncated by interaction with the host cluster. 
\end{abstract}




\section{Introduction} \label{intro}

The circumgalactic medium (CGM) is a gas reservoir
 surrounding galaxies within their dark matter haloes.
Recent cosmological simulations and observations support that 
 the CGM is closely connected to the evolution of galaxy
 via gas flows \citep[see][for a review]{tum17}.
In the sense that the CGM harbours
 metal-enriched gas ejected from the galaxy, driven by star formation feedback, 
 and fresh gas accreted from the intergalactic medium (IGM) 
 to fuel star formation,
 the CGM provides an excellent laboratory 
 for better understanding the mechanisms that control galaxy evolution.

It is challenging to detect the emission from CGM 
 with currently available facilities because of the low surface brightness,
 although there are several successful attempts from 
 deep imaging and spectroscopic observations \citep[e.g.][]{can14,zha16,dad20}. 
Therefore, Ly$\alpha$ and metal absorption lines imprinted in
 background source spectra have been widely used to trace the CGM
 \citep[e.g.][]{ade05,ste10,pro13}.
For the background source, quasars are popular
 because they are luminous objects with a simple spectral shape, 
 generally characterized by a power-law continuum and broad emission lines,
 enabling us to detect foreground absorbers 
 with a high signal-to-noise ratio (S/N).    
The Ly$\alpha$ absorption line is very sensitive to the existence of CGM,
 but does not allow us to construct a large sample of Ly$\alpha$ absorbers
 due to observational limitations \citep[but see][]{leh18}.
Among the metal absorption lines,
 the \MgII\ $\lambda\lambda$2796, 2803 doublet is particularly useful
 for statistical studies 
 because it is a prominent feature, visible from the ground,
 spanning a wide redshift range ($0.3<z<2.5$) in the optical wavelengths.
The \MgII\ absorption is known to probe gas 
 at temperature $\sim10^4$ K \citep{cha03} 
 with the neutral hydrogen column density 
 $\approx~10^{16}$--$10^{22}$ cm$^{-2}$ \citep{rao06}.

Since the first discovery of an intervening \MgII\ absorption 
 from quasar-galaxy pairs by \citet{ber86},
 the relation between galaxies and their CGM has been investigated
 over the last three decades.
The strength of \MgII\ absorption at fixed impact parameter is found to 
 be correlated with galaxy properties such as luminosity, mass, color and
 (specific) star formation rate 
 \citep[e.g.][]{bou06,che10a,bor11,lan14,rub18},  
 indicating that massive star-forming galaxies 
 have a relatively dense and/or extended CGM
 compared to low-mass passive galaxies.
The CGM is also related to galaxy orientation \citep[e.g.][]{ste02,kac12},
 originating from that gas accretion on to galaxies is preferentially 
 along their major axis, whereas gas outflows along the minor axis.

The CGM is expected to be affected by galaxy environment 
 as well as galaxy properties.
The \ion{H}{i} observations in the local Universe 
 reveal that gas around galaxies has complex structures 
 when the galaxy interacts with neighbour galaxies or
 the host group/cluster \citep[e.g.][]{chy08,chu09,wol13}.
The combined analysis of internal velocity dispersion of galaxies 
 and strong lensing predicts that 
 the halo size of cluster galaxies is significantly 
 different from that of field galaxies \citep[e.g.][]{mon15,mon17}.
However, the results from previous studies with the \MgII\ doublet 
 do not seem to agree. 
For example, \citet{lop08} and \citet{pad09} found that 
 strong \MgII\ absorbers are abundant in cluster central regions, 
 while the numbers of weak absorbers are compatible 
 between cluster and field regions.
The authors suggested that many galaxies are in clusters,
 but their haloes are truncated by cluster environment effects.
On the other hand, \citet{che10b} reported that 
 the strengths of \MgII\ absorption in group environments 
 are similar to those in isolated environments.
\citet{bor11} pointed out that the extended \MgII\ gas 
 around group galaxies could be explained 
 by a superposition of member galaxies \citep[but see][]{nie18},
 implying that the group environment does not have a significant impact 
 on the CGM of individual galaxies.

To systematically study how the CGM is affected by the cluster environment, 
 we construct a large sample of quasar-cluster pairs 
 using data sets from the Sloan Digital Sky Survey \citep[SDSS;][]{yor00}
 and detect \MgII\ absorption lines in the quasar spectra. 
The outline of this paper is as follows.
We explain the data used, including cluster, quasar and galaxy samples, 
 in Section \ref{data}.
We describe the methods to measure the equivalent widths (EWs) 
 of \MgII\ lines and to select \MgII\ absorbers in Section \ref{anal}.
We present and discuss our findings in Section \ref{result}.
We summarize the results in Section \ref{summ}.
Throughout, we assume a flat $\Lambda$CDM cosmology with
 $H_0$ = 100 $h$ \kms\ Mpc$^{-1}$,
 $\Omega_{\Lambda}$ = 0.7 and $\Omega_m$ = 0.3.
We omit the terms of `$-$5 log $h$' and `$h^{-1}$'
 from absolute magnitude and cluster mass units, respectively.
All magnitudes are given in the AB system and 
 all measurement errors are 1$\sigma$.

\section{Data} \label{data}

\subsection{Galaxy cluster sample} \label{cluster}

We use the \redm\
 \citep[red-sequence Matched-filter Probabilistic Percolation;][]{ryk14,ryk16}
 catalogue to search for \MgII\ absorbers associated with galaxy clusters.
The \redm\ is one of the largest and homogeneous samples of galaxy clusters.
This catalogue is constructed with sophisticated algorithms
 by sampling red-sequence galaxies as potential cluster members
 based on a minimal spectroscopic training set.
The \redm\ richness parameter (\redl) is well correlated with 
 other proxies of cluster mass
 and about 90 per cent of member candidates are spectroscopically confirmed
 to be in the host cluster even at the low richness
 \citep[e.g.][]{rin18,soh18}.
Therefore, although the \redm\ clusters are photometrically selected,
 most of them are expected to be genuine clusters.

In this study, 
 we adopt the version 6.3 of \redm\ 
 catalogue\footnote{http://risa.stanford.edu/redmapper/}
 containing $\sim$26 000 galaxy clusters in the SDSS 
 Data Release 8 \citep[DR8;][]{aih11}.
The dynamical mass ($M_{200}$) and projected velocity dispersion (\reds) 
 of \redm\ clusters are estimated from \redl\
 using the scaling relations in table 4 of \citet{rin18}:
 \begin{align}
 {\rm log}(M_{200}/10^{14.5}\ {\rm M}_{\sun})=
 0.580\ {\rm log}(\lambda_{\rm red}/100)+0.005,\\
 {\rm log}(\sigma_{\rm red}/700\ {\rm km\ s^{-1}})=
 0.240\ {\rm log}(\lambda_{\rm red}/100)+0.046.
 \end{align}
Then, their virial radius ($r_{200}$) is converted from \reds\ 
 with the formula of \citet{car97}:
 \begin{align}
 r_{200}/{\rm Mpc}=\frac{\sqrt{3}\ \sigma_{\rm red}}{10\ H(z_{\rm red})},
 \end{align}
 where \redz\ is the \redm\ cluster redshift
 from the photometric redshifts of member candidates
 and $H$(\redz) is the Hubble parameter at \redz\ \citep{pee93}.

\begin{figure}
\includegraphics[width=\columnwidth]{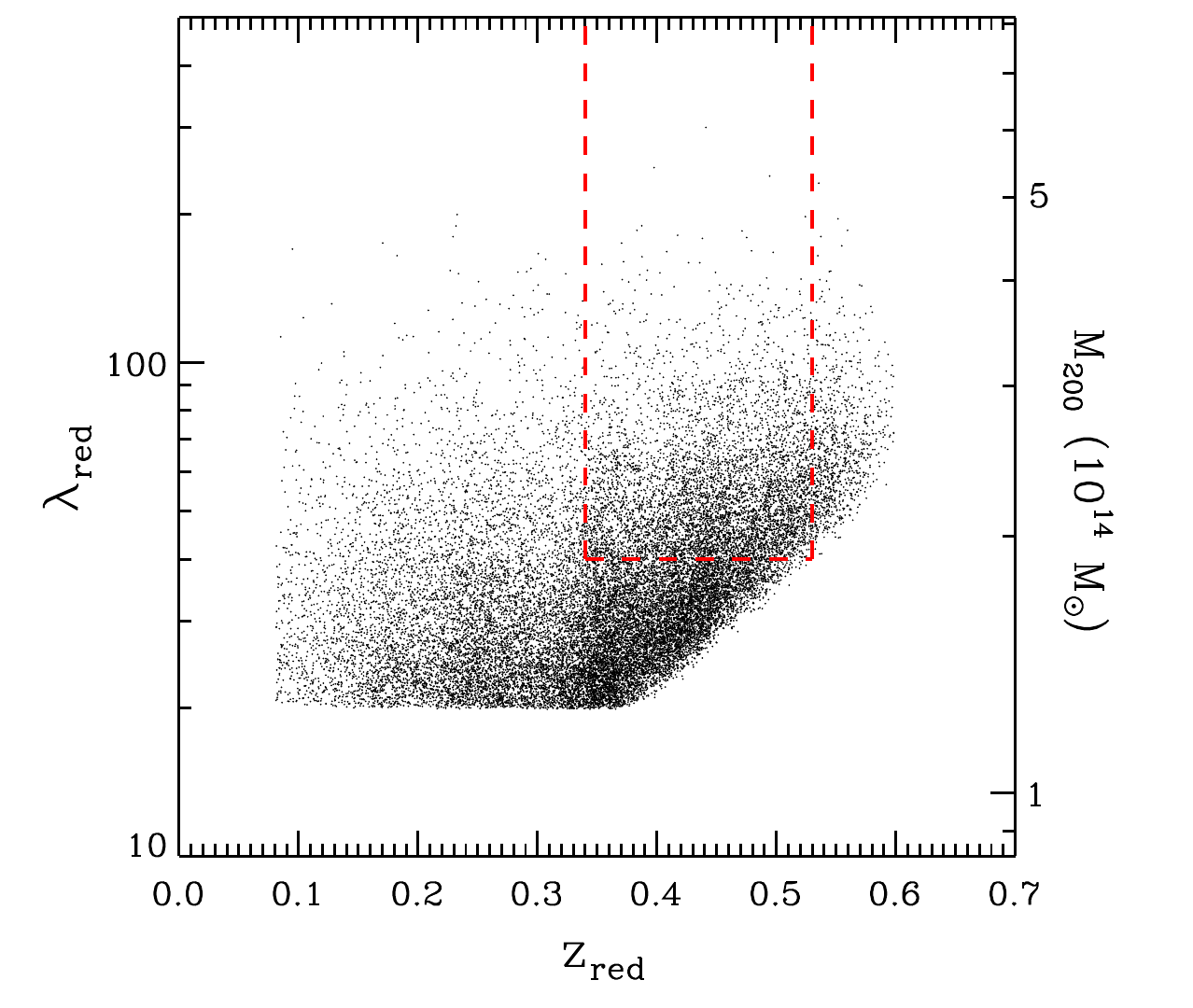}
\caption{Richness parameter versus redshift diagram of \redm\ 
galaxy clusters. Red dashed lines denote our selection criteria.
\label{fig-sample}}
\end{figure}

Figure \ref{fig-sample} displays the \redl\ and \redz\ distributions.
We focus on 6179 clusters, surrounded by red dashed lines, 
 at \redz\ = 0.34--0.53 and \redl\ $\gid$ 40,
 corresponding to $M_{200} \gid 1.88 \times 10^{14}~{\rm M}_{\sun}$.
The lower limit of \redl\ is necessary to guarantee 
 the scaling relations of \citet{rin18}:
 their figures 9 and 11 show that the relations are not reliable 
 at \redl\ $<$ 40.
The upper limit of \redz\ is for making a volume-limited sample
 at \redl\ $\gid$ 40.
The lower limit of \redz\ is determined to ensure that 
 the \MgII\ doublet is comfortably within the BOSS 
 (Baryon Oscillation Spectroscopic Survey; \citealt{daw13}) spectral coverage 
 (3650--10 400 \AA)\footnote{When using quasar spectra obtained with 
 the SDSS spectrograph (3800--9200 \AA), the redshift cut of \redz\ $\gid$ 0.34
 may not be enough to securely measure the \MgII\ doublet. 
 If we remove the SDSS spectrograph data or strictly narrow the redshift range 
 to \redz\ = 0.39--0.53, the number of quasar-cluster pairs 
 is reduced by 14 or 26 per cent, respectively. 
 This does not change the main results of this study,
 but makes our conclusion less significant.}  
 at \redz$-$5\reds(1+\redz)/$c < z <$ \redz$+$5\reds(1+\redz)/$c$, 
 where $c$ is the speed of light,
 which is the redshift range where we search for absorbers 
 ($\Delta v$/\reds\ $<$ 5; the search window hereafter). 

\subsection{Quasar sample} \label{qso}

We use the quasar catalogue of \citet{par18}, 
 which is from the SDSS DR14 \citep{abo18} 
 and contains $\sim$526 000 quasars.
Among these, 15 and 85 per cent were observed with 
 SDSS and BOSS spectrographs, respectively.
Their optical spectra are downloadable from the SDSS Science Archive 
 Server\footnote{http://dr16.sdss.org/sas/dr16/sdss(eboss)/spectro/redux/}.

\begin{figure}
\includegraphics[width=\columnwidth]{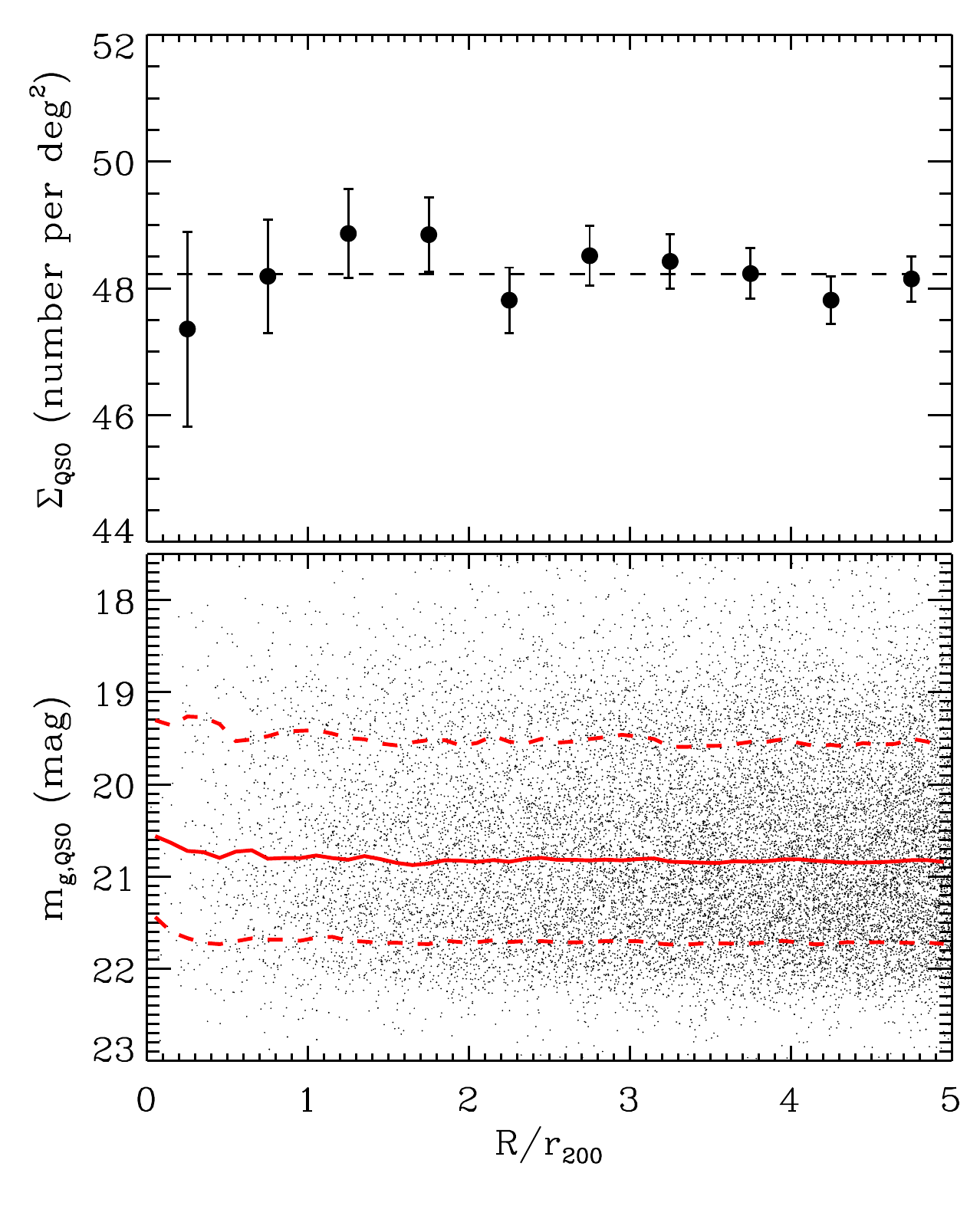}
\caption{Number density (top) and $g$-band magnitudes (bottom) of quasars
cross-matched with \redm\ clusters
as a function of the cluster-centric radius normalized to its virial radius.
In the top panel, the horizontal dashed line is the mean number density
and error bars are calculated from the binomial statistics.
In the bottom panel, the red solid line indicates the sliding median and 
the dashed lines enclose 68 per cent of quasars.
For clarity, (randomly selected) 20 per cent of quasars are presented.
\label{fig-distq}}
\end{figure}

By cross-matching the quasar and cluster catalogues, 
 we search for quasars with projected distance 
 from the centre of each cluster centre smaller than 5$r_{200}$.
This distance is large enough to cover both cluster to field environments 
 so that we can examine any difference between the two environments.
We find 96 759 quasars (including duplicates) in and around the clusters and 
 hereafter call them quasar-cluster pairs for convenience. 
There is little difference in the redshift distribution 
 of quasars (\zqso) between parent and matched samples. 
In Figure \ref{fig-distq}, we present the number density and 
 $g$-band point spread function magnitudes of the background quasars 
 as a function of the cluster-centric radius normalized by $r_{200}$ 
 ($R/r_{200}$).
The mean number density is 48.23$\pm$0.16 per deg$^{2}$ and
 typical range (68 per cent enclosure) of $g$-band magnitudes is 
 19.54--21.71 mag.
Because the quasars and clusters are selected independently from each other,
 the quasar properties show no systematic variation 
 with the cluster-centric radius.

\subsection{Galaxy sample} \label{galaxy}

\begin{figure*}
\includegraphics[width=\linewidth]{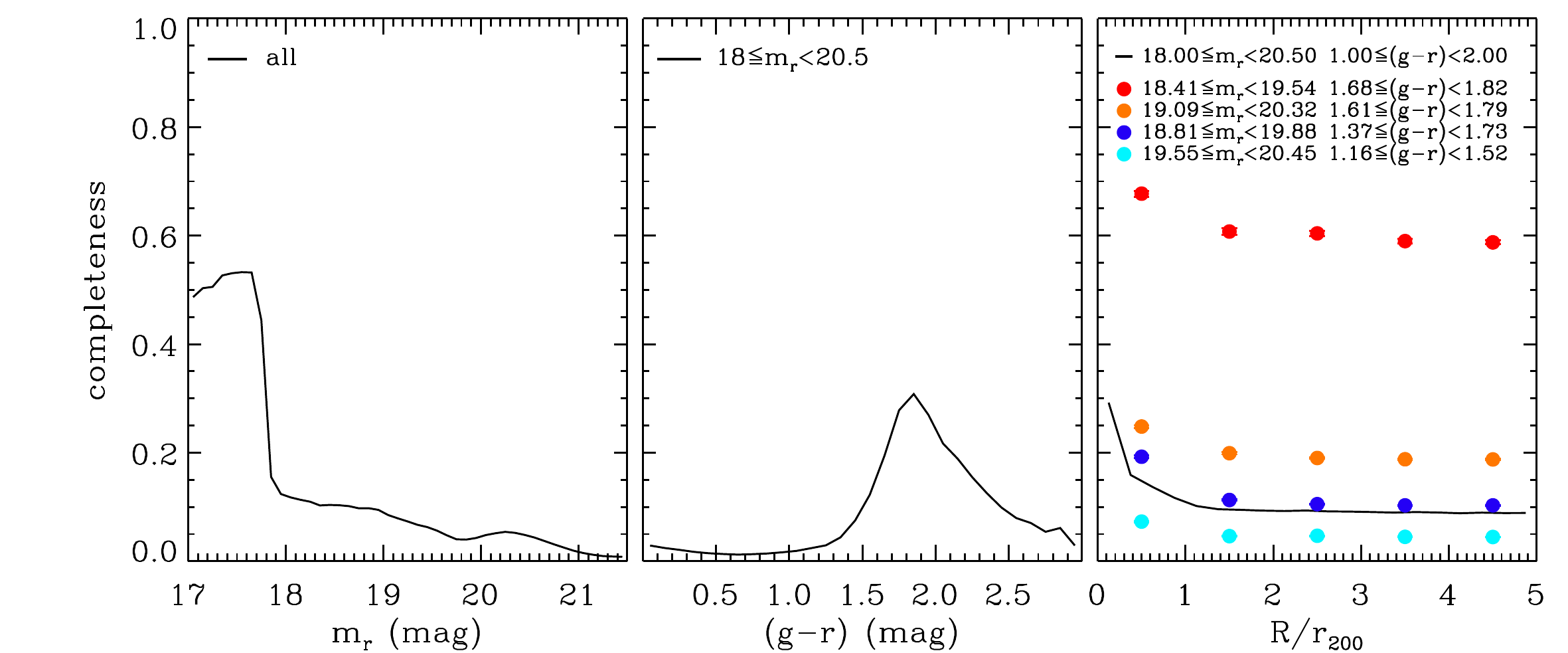}
\caption{Completeness of the spectroscopic sample of galaxies 
in and around \redm\ clusters as functions of
$r$-band Petrosian magnitude (left), $(g-r)$ model colour (middle) 
and cluster-centric radius (right).
In the middle panel, galaxies with $m_{r}$ = 18.0--20.5 are used 
to unveil the colour dependence of completeness
in galaxies within our search window.
In the right panel, the $m_{r}$ and $(g-r)$ cuts are denoted for each case.
The solid line is for all galaxies within the search window.
The red, orange, purple and cyan circles are for 
the bright-red, faint-red, bright-blue and faint-blue subsamples, respectively.
\label{fig-compn}}
\end{figure*}

We also compile galaxy data 
 using a photometric sample of galaxies in the SDSS DR16 \citep{ahu20}.
Their redshift information is obtained from the SDSS database 
 and is supplemented from literature \citep[see][for details]{hwa10,hwa14}.
At $R/r_{200} <$ 5, the numbers of galaxies 
 in the photometric and spectroscopic samples are 
 about 12.26 million and 486 000, respectively.
The spectroscopic completeness as a function of 
 $r$-band Petrosian magnitude ($m_r$) is shown 
 in the left panel of Figure \ref{fig-compn}.
The limiting magnitude of SDSS main galaxy survey is 
 $m_{r} =$ 17.77 \citep{str02}. 
However, the completeness is not small even at $m_{r} >$ 17.77
 because of other subprograms targeting faint extra-galaxies   
 such as the Luminous Red Galaxy \citep[LRG;][]{eis01},
 Emission Line Galaxy \citep[ELG;][]{com16},
 SPectroscopic IDentification of ERosita Sources \citep[SPIDERS;][]{cle16} and
 Extended Baryon Oscillation Spectroscopic Survey \citep[eBOSS;][]{daw16}.
We note that a significant fraction of \redm\ clusters 
 were not covered by the main galaxy survey, 
 which makes the completeness for the cluster regions 
 reaches only half even at $m_{r} <$ 17.77.

Among $\sim$486 000 galaxies with redshift,
 64 716 galaxies are found to be within the search window.
The primary source of them is the LRG sample (79 per cent)
 rather than the main galaxy sample (10 per cent) because of redshift range.
Their $m_r$ and $(g-r)$ model colour ranges are
 about 18.0--20.5 and 1.0--2.0 mag, respectively.
When fixing $m_r$ from 18.0 to 20.5 mag,
 we can see the colour dependence of spectroscopic completeness
 as shown in the middle panel of Figure \ref{fig-compn}:
 high completeness in the red colour range
 connected to the selection criteria for LRGs.
At $m_{r} <$ 17.77, dominated by the main galaxy sample,
 this colour dependence disappears
 because there is no colour selection for the main galaxy sample.
At $m_{r}$ = 18.0--20.5 and $(g-r)$ = 1.0--2.0,
 we can perceive the radial dependence of completeness
 (the solid line in the right panel) as well.
The high completeness in the cluster region 
 may originate mainly from that the \redm\ cluster finding technique
 prefers galaxies with redshift as member galaxies,
 especially for the central galaxy (see section 8 of \citealt{ryk14}),
 and partly from that some \redm\ cluster galaxies are targeted
 in the SPIDERS programs. 

\begin{figure}
\includegraphics[width=\columnwidth]{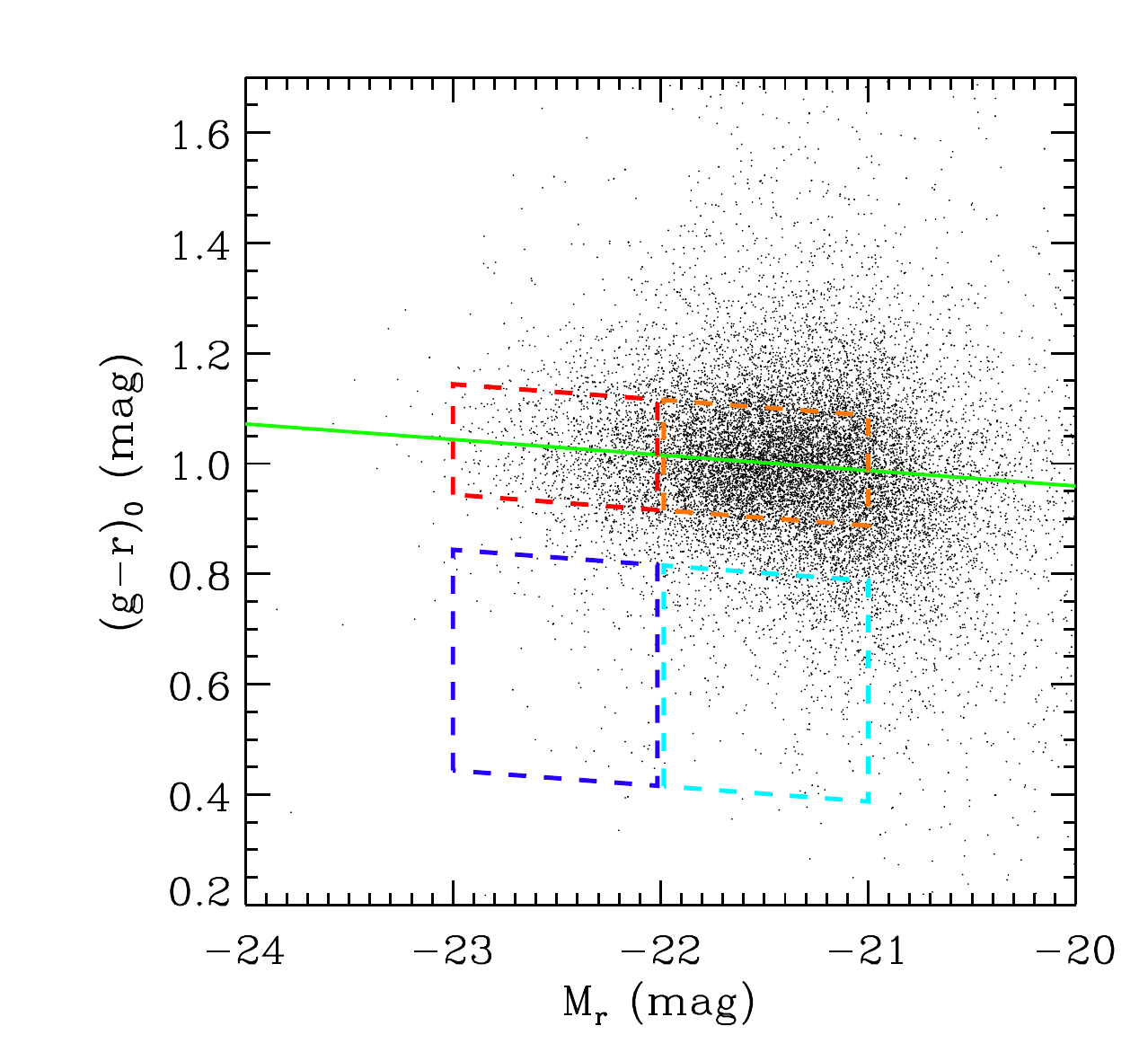}
\caption{Colour-magnitude diagram of galaxies 
associated with the \redm\ clusters.
The green line stands for the red sequence.
The red, orange, purple and cyan boxes represent the selection criteria of
bright-red, faint-red, bright-blue and faint-blue galaxies, respectively.
For clarity, (randomly selected) 30 per cent of galaxies are presented.
\label{fig-cmd}}
\end{figure}

We regard 64 716 galaxies within the search window
 as galaxies associated with an ensemble of \redm\ clusters
 (the \redm\ cluster galaxies hereafter)
 and plot their colour-magnitude diagram in Figure \ref{fig-cmd}.
The absolute magnitude $M_{r}$ and $(g-r)_{0}$ colour are obtained 
 after $K$-correction \citep{bla07} and evolution correction \citep{teg04}.
Because most of them are LRGs,
 the so-called red sequence is well noticeable, whereas the blue cloud is not.
Using galaxies brighter than $M_{r}$ = $-$21, 
 at which the number of galaxies per mag is peaked,
 we fit the red sequence to a linear relation (green line):
 \begin{equation}
 (g-r)_{0} = -0.028\ M_{r}+0.397\ (1\sigma\ {\rm scatter} = 0.107).
 \end{equation}
We then define the four subsamples of cluster galaxies 
 according to their $M_{r}$ and colour deviation from the sequence 
 $\Delta(g-r)_{0}$, which are\\
\begin{tabular}{r}
     bright-red: $-23\lid M_{r}<-22$ \& $-0.1\lid \Delta(g-r)_{0}<+0.1$,\\
      faint-red: $-22\lid M_{r}<-21$ \& $-0.1\lid \Delta(g-r)_{0}<+0.1$,\\
    bright-blue: $-23\lid M_{r}<-22$ \& $-0.6\lid \Delta(g-r)_{0}<-0.2$,\\
 and faint-blue: $-22\lid M_{r}<-21$ \& $-0.6\lid \Delta(g-r)_{0}<-0.2$.\\
\end{tabular}
These subsamples include 4919, 24 615, 395 and 2523 galaxies. 
The typical stellar masses\footnote{The stellar mass and star formation rate
 are drawn from the MPA-JHU value-added galaxy catalogues 
 \citep[][]{gal05,sal07}.}
 of the bright and faint galaxies are 4.4--7.8 $\times 10^{11}$ and 
 1.4--4.0 $\times 10^{11}~{\rm M}_{\sun}$, respectively. 
The specific star formation rates of the red and blue galaxies are 
 10$^{-12.4}$--10$^{-11.4}$ and 10$^{-10.4}$--10$^{-9.6}$ yr$^{-1}$.
Their apparent $m_{r}$ and $(g-r)$ ranges are denoted 
 in the right panel of Figure \ref{fig-compn}.
The overall completeness in $R/r_{200}$ bins
 for the bright-red, faint-red, bright-blue and faint-blue galaxies is also
 presented with red, orange, purple and cyan circles, respectively.
The completeness differs between the subsamples 
 (high completeness in the bright/red galaxies),
 but has a similar radial trend (higher completeness in the inner region).

\section{Analysis} \label{anal}

\begin{figure*}
\includegraphics[width=0.85\linewidth]{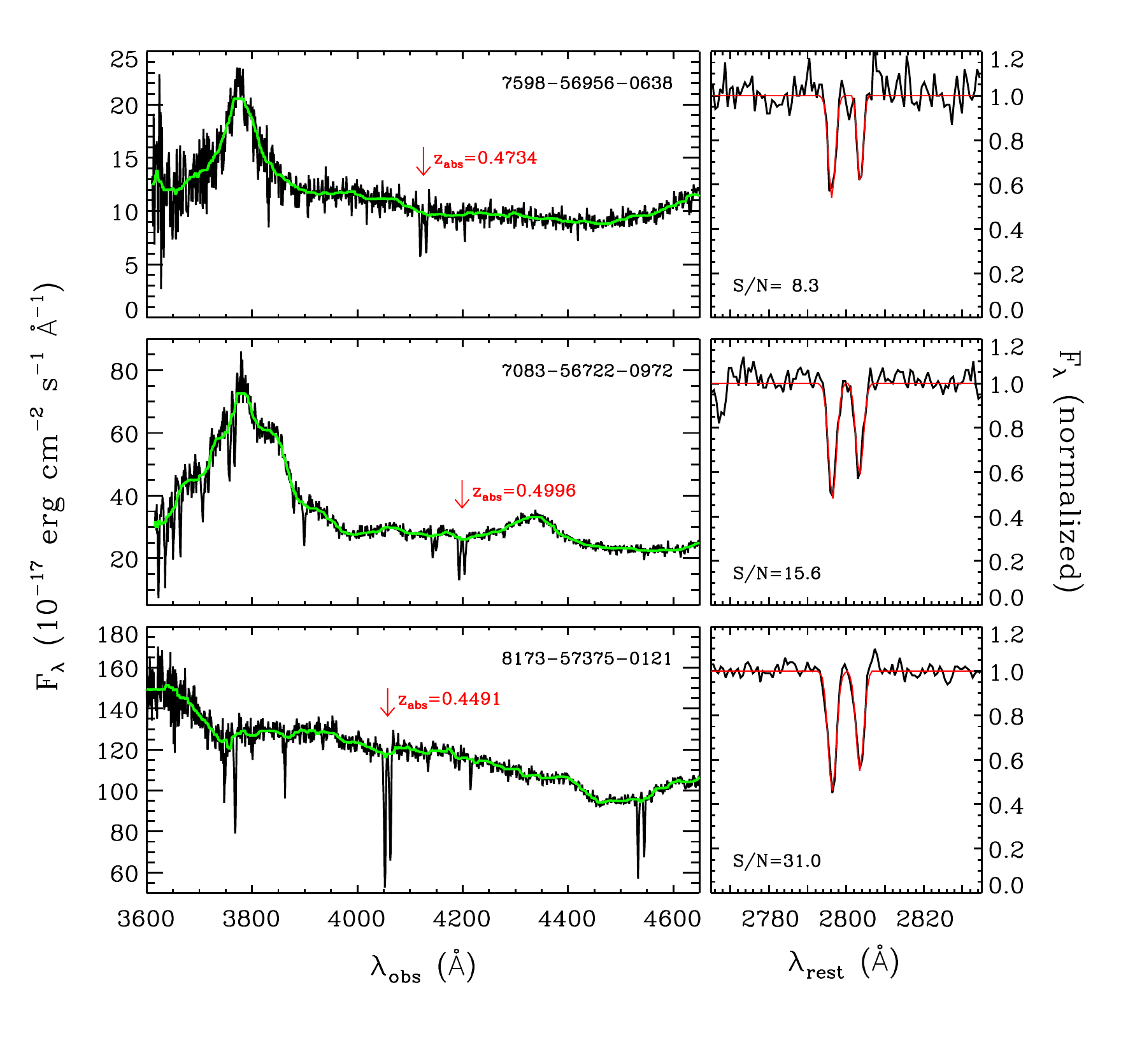}
\caption{Examples of \MgII\ doublet detected spectra. 
In the left panels, the observed spectrum (ID denoted in the top-right corner)
and median-smoothed global continuum are in black and green, respectively.
We mark the redshift at which the \MgII\ absorber is found 
with a downward arrow. 
In the right panels, the (rest-frame) continuum-normalized spectrum and 
the two-Gaussian model fitting result are in black and red, respectively.
The S/N of total EW of \MgII\ lines is presented 
in the bottom-left corner.
\label{fig-exam}}
\end{figure*}

To detect \MgII\ absorption features in the quasar spectrum, 
 we first estimate the quasar continuum.
After masking bad pixels flagged by the SDSS pipeline,
 we use a simple smoothing method with a median filter.
The filter size is changed from 15 to 80 times 
 the instrumental resolution 
 ($\sim$1.0 \AA; $\sim$70 \kms\ for both SDSS and BOSS spectrographs)
 according to the S/N at each pixel
 (i.e. 15 times at S/N $>$ 100 and 80 times at S/N $<$ 5),
 resulting in an adaptive smoothing effect.
This is determined empirically to be small enough so that 
 the global spectral shape and broad emission lines
 of background quasar should be included in the continuum estimation
 and to be large enough so that narrow line features
 such as the \MgII\ doublet from foreground absorbers should not be included.
The left panels of Figure \ref{fig-exam} show three examples of 
 the quasar spectra along with the continuum fitting result.

Within the search window, we attempt to measure the EWs
 of \MgII\ doublet lines (\ewa\ \& \ewb) 
 in every step of $\Delta z = 5\times10^{-5}$ 
 ($\sim$1/7 of instrumental resolution).
The continuum-normalized observed spectrum 
 is transformed to the rest-frame at a potential redshift,
 and then renormalized by the local continuum
 (determined  from a linear fitting around the doublet 
 at 2765--2785 and 2815--2835 \AA) to remove the remaining residual.
Using the {\small MPFIT/IDL} package \citep{mar09},
 the \MgII\ doublet is modelled with two Gaussians,
 as shown in the right panels of Figure \ref{fig-exam}.
The wavelength separation of the two Gaussians 
 (usually 7.2 \AA\ for the \MgII\ doublet) 
 is allowed to vary with a tolerance of $\pm$2 times 
 the (rest-frame) resolution.
The Gaussians have the same linewidth, which is limited 
 from 90 per cent of the resolution
 to 3.1 \AA\footnote{By considering the uncertainty of 
 instrumental resolution information provided by the pipeline, 
 we regard the 90 per cent value as the minimum width of \MgII\ lines.
 When adopting the 100 per cent value as the lower limit, 
 we lose about 20 per cent of absorbers.     
 If the full width at half maximum (FWHM) of \MgII\ lines  
 is larger than the separation, 
 it is difficult to identify the doublet based on our scheme.
 We thus focus only on narrow 
 (velocity dispersion \disps\ $<$ 200 \kms) 
 absorption line systems in this study.}.
Because the line ratio of the doublet \ratio\ is theoretically bounded 
 between 2 (unsaturated) and 1 (completely saturated), 
 we keep the measurement with
 $1-\sigma_{\rm ratio} \lid$ \ratio\ $\lid 2+\sigma_{\rm ratio}$, 
 where $\sigma_{\rm ratio}$ is the ratio error
 calculated using the uncertainties in Gaussian parameters.
The S/N $\gid$ 3 is taken as the detection threshold 
 for both \ewa\ and \ewb.
There is another condition that the height of each Gaussian is 
 three times larger than the flux fluctuation (root mean square) 
 around the \MgII\ doublet, which means that the doublet should be 
 a major line feature at least around it,
 helpful for preventing false detections (see also next paragraph).
More than one \MgII\ absorber could be identified in a given spectrum.
Therefore, as we measure EWs by increasing the redshift for each step, 
 we keep all the redshifts that give local maxima of the S/N for total EW 
 as the absorber redshifts (\zabs). 
This automated procedure results 362 possible \MgII\ absorber candidates
 from 82 360 quasar spectra.

\begin{figure}
\includegraphics[width=\columnwidth]{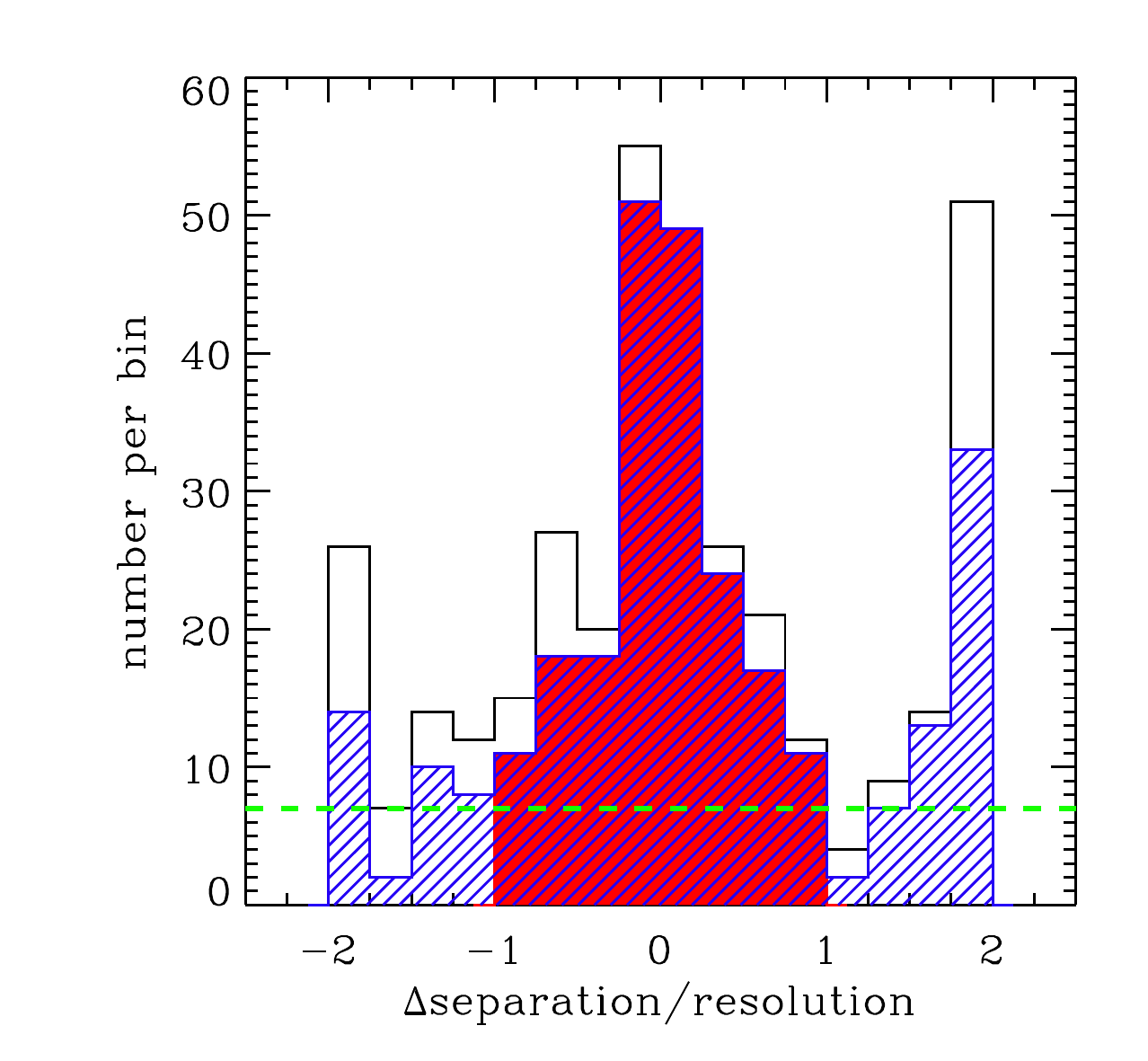}
\caption{Distribution of the difference between the measured wavelength 
separation and the intrinsic value of 7.2 \AA\ 
divided by the instrumental resolution.
The black solid, blue hatched and red shaded histograms
are for the 362 possible candidates from the automated process,
288 probable candidates through our visual inspection, 
and 197 \MgII\ absorbers finally selected, respectively.
The green dashed line indicates the estimated contamination level per bin.
\label{fig-sepa}}
\end{figure}

Because a substantial fraction of the 362 possible candidates 
 can be contaminated by false detections, 
 we visually inspect them to select more reliable ones.  
We exclude the measurements with the Gaussian fits 
 not matched to the actual line 
 profiles\footnote{It is difficult to remove them 
 simply by using a $\chi^{2}$ threshold 
 because there could be adjacent line features.}. 
The measurements based on inaccurate continuum 
 estimates\footnote{Our continuum estimation is not satisfactory 
 when there are broad absorption lines or 
 many adjacent (narrow) absorption lines.
 It is also incorrect near the peak (not the wing) of broad emission lines.
 To handle these cases properly, more detailed modelling is necessary.}
 are also eliminated.
As a result, 288 probable candidates remain.
In Figure \ref{fig-sepa}, we present the histogram of the values obtained 
 by subtracting 7.2 \AA\ from the wavelength separation 
 between the two Gaussians and dividing by the instrumental resolution
 ($\Delta$separation/resolution).
A significant fraction of our candidates are genuine 
 in the sense that the $\Delta$separation/resolution distribution 
 is concentrated around zero.
If we do not adopt the criterion that the Gaussian heights 
 are three times larger than the flux fluctuation,
 this concentration becomes much weaker.
On the other hand, this distribution has secondary peaks at $\pm$2
 because of the measurements enforced by the boundary condition.
We finally select 197 \MgII\ absorbers
 with abs($\Delta$separation/resolution) $\lid$ 1
 to further reduce contamination.
The number of contaminants in the final list is estimated to be 56 
 by considering the number of probable candidates with
 abs($\Delta$separation/resolution) = 1--1.75 (seven absorbers per bin).
Therefore, we expect that the reliability of our identification 
 is (1$-$56/197) $\sim$72 per cent.

We note that all the 197 \MgII\ absorbers have redshifts 
 smaller than those of quasars (i.e. \zabs\ $<$ \zqso). 
Among them, there are four absorbers with \zqso$-$\zabs\ $<$ 0.003, 
 corresponding to the relative velocity of $\Delta v <$ 600 \kms.
Although these absorbers are potentially associated with their quasars 
 (i.e. not intervening absorbers), 
 they are kept in the final list for consistency. 
On the other hand, there are five absorbers matched with two different clusters. 
We thus have 202 Mg II absorber-cluster pairs. 
In a given single spectrum of quasar, 
 there could be multiple absorbers at different redshifts. 
However, we find no such case in this study due to the low detection rate 
 (only 0.6 per cent even in cluster regions; see Section \ref{radial}).

\section{Results and Discussion} \label{result}

\subsection{Absorber properties} \label{absorber}

\begin{figure*}
\includegraphics[width=0.9\linewidth]{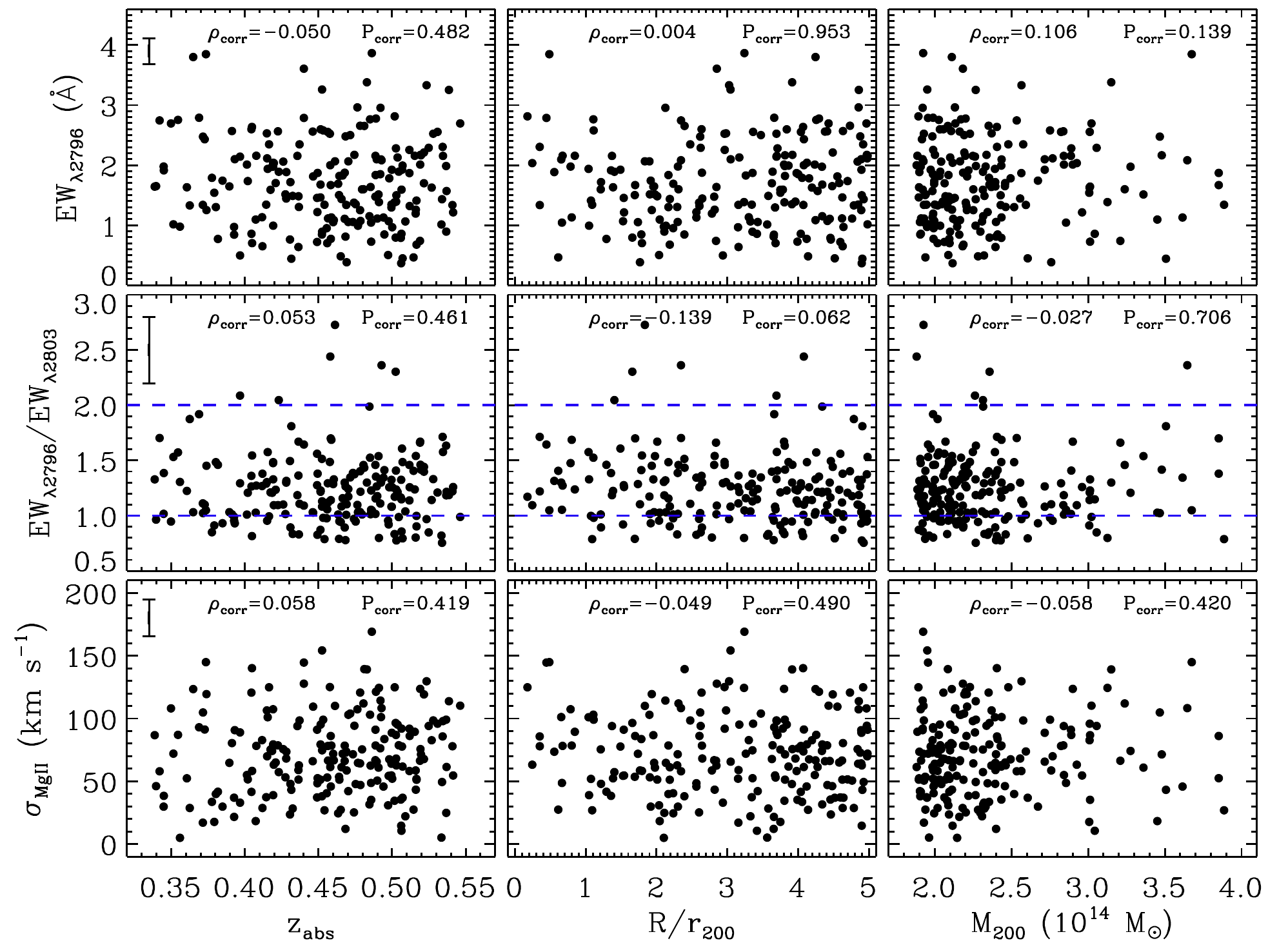}
\caption{
\MgII\ absorber properties as functions of absorber redshift (left), 
cluster-centric radius (middle) and cluster mass (right). 
The EW of \MgII\ $\lambda$2796, doublet line ratio and velocity dispersion of \MgII\ lines 
are presented in the top, middle and bottom panels, respectively. 
Their median errors are denoted in the top-left corner. 
The numbers in each panel indicate the Spearman rank correlation coefficient $\rho_{\rm corr}$ 
and the probability of obtaining the correlation by chance P$_{\rm corr}$. 
The blue horizontal dashed lines are the theoretical limits of the line ratio.
\label{fig-abs}}
\end{figure*}

We provide the basic information of 197 \MgII\ absorbers in Table \ref{tab} and 
 plot the measured parameters as functions of \zabs, $R/r_{200}$ and $M_{200}$ 
 in Figure \ref{fig-abs}.
The sensitivity for detecting \MgII\ absorbers 
 gradually increases with redshift until $z \approx$ 1.5 
 because the SDSS spectroscopic throughput reaches a maximum at $\sim$7000 \AA. 
However, we find no correlation between \ewa\ and \zabs\ 
 in the sense that the probability of obtaining the given correlation by chance 
 in the Spearman test, shown in the top-left panel, is not small enough. 
This suggests that there is no significant selection bias within our redshift range. 
We also note that the measured parameters are correlated neither with 
 $R/r_{200}$ nor $M_{200}$; 
 this dependence does not change even if we add the sample of \citet{lop08}.
There are publicly available catalogues of \MgII\ absorbers, 
 which are selected regardless of their environment \citep[e.g.][]{zhu13,rag16}.
The 36 absorbers are also included in the catalogue of \citet{zhu13}.
For these common absorbers,
 two measurements are consistent with each other
 (see Appendix \ref{appen} for more details).  

\begin{table*}
\centering
\caption{Basic information of 197 \MgII\ absorbers. 
Columns 1--4 are from the quasar catalogue of \citet{par18}. 
Columns 5--8 are our measurements and errors.
The absorber redshift errors are typically 0.0003.
The full table is available online.}
\label{tab}
\begin{tabular}{cccccccc}
\hline
spectrum ID     &  RA        & Dec        & \zqso\ & \zabs\    & \ewa\          & \ratio\        & \disps\         \\
                &  deg.      & deg.       &        &           & \AA\           &                & \kms\           \\
\hline
7830-57043-0126 & 034.956963 &  01.454112 & 1.4497 &    0.3736 & 3.85 $\pm$0.63 & 1.05 $\pm$0.27 & 144.9 $\pm$26.8 \\
6468-56311-0076 & 151.811924 &  31.772667 & 3.3370 &    0.3716 & 2.48 $\pm$0.31 & 1.02 $\pm$0.17 & 104.8 $\pm$16.5 \\
4608-55973-0567 & 133.845293 &  37.907218 & 2.3020 &    0.4074 & 1.09 $\pm$0.14 & 1.03 $\pm$0.18 & 018.3 $\pm$11.2 \\
4572-55622-0204 & 146.842194 &  38.014749 & 2.4010 &    0.4584 & 1.67 $\pm$0.22 & 1.70 $\pm$0.39 & 086.1 $\pm$14.6 \\
4572-55622-0156 & 146.997598 &  38.379074 & 2.2780 &    0.4494 & 1.87 $\pm$0.25 & 1.38 $\pm$0.30 & 052.4 $\pm$11.9 \\
\hline
\end{tabular}
\end{table*}

\begin{figure*}
\includegraphics[width=0.9\linewidth]{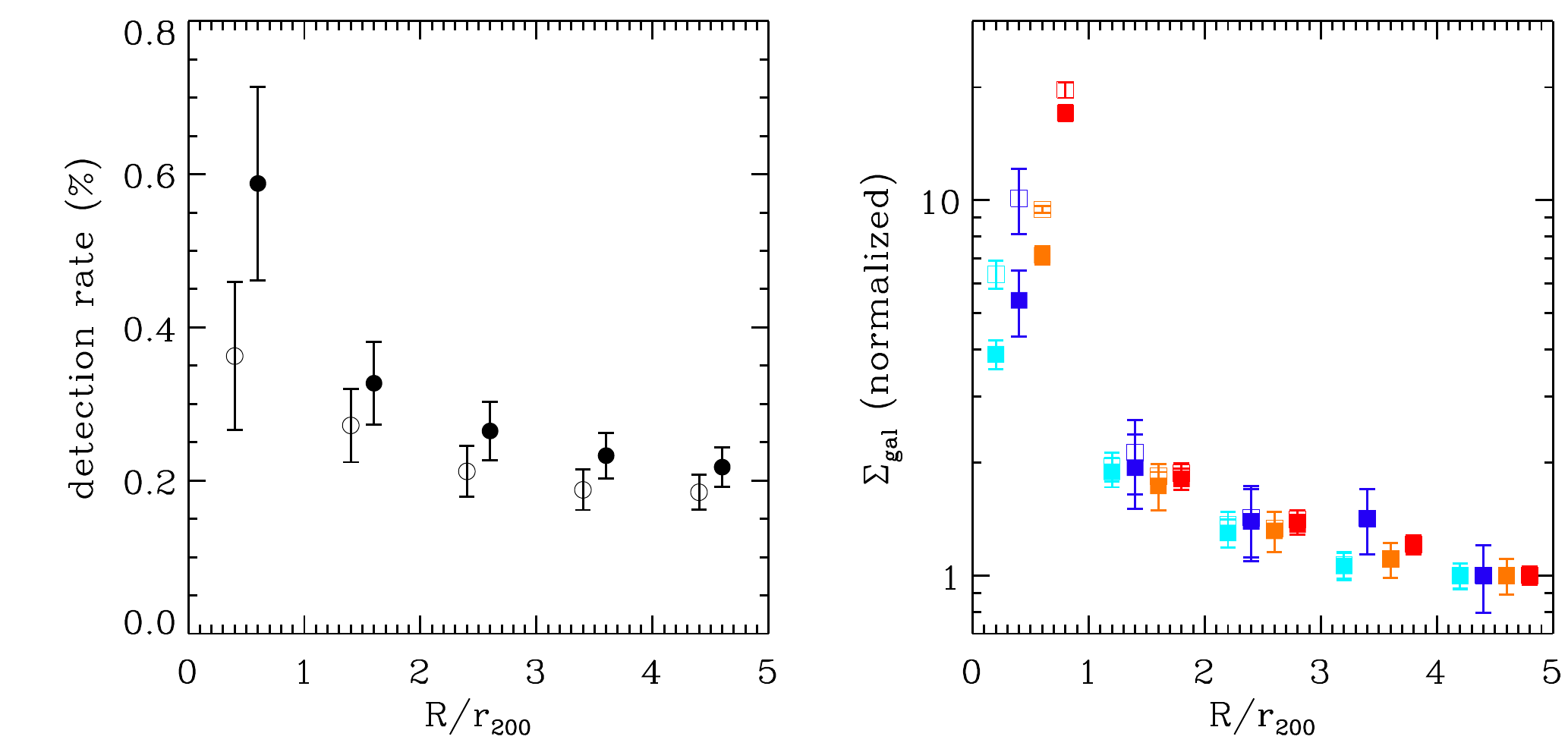}
\caption{\MgII\ absorber detection rate per quasar (left)
and the number density of four subsamples of cluster galaxies,
normalized by the value at $R/r_{200} >$ 4 (right), 
as a function of cluster-centric radius.
In the right panel,
the red, orange, blue and cyan squares are for
the bright-red, faint-red, bright-blue and faint-blue galaxies,
respectively.
The open and filled circles/squares indicate
the raw and corrected values.
These symbols are properly separated for better visibility.
\label{fig-rate}}
\end{figure*}

The typical ranges of \ewa, \ratio\ and \disps\footnote{We 
 obtain the intrinsic linewidth by subtracting the 90 per cent of 
 instrumental resolution in quadrature.} 
 are 0.97--2.52 \AA, 0.95--1.49 and 37--105 \kms, respectively.
The \MgII\ absorbers are often classified into weak and strong 
 systems using the division value of 
 \ewa\ = 0.3 \AA\ \citep[e.g.][]{lop08}.
If we follow this classification scheme, 
 all our absorbers are strong systems. 
The eight absorbers are even ultra-strong (\ewa\ $>$ 3 \AA).
We note that, based on the completeness test result of \citet{zhu13}, 
 more than 60 per cent of \MgII\ absorbers with \ewa\ $>$ 1.0 \AA\  
 could be detected around z = 0.45
 from the SDSS quasar spectra, but less than 10 per cent for \ewa\ $<$ 0.3 \AA.
The \ratio\ distribution is peaked at $\sim$1.2, 
 indicating that most of the doublets are saturated. 
About 80 per cent of absorbers have \disps\ $<$ 100 \kms,
 which are compatible with gas velocity dispersion around ELGs 
 rather than LRGs \citep[][]{lan18}.
These absorber properties are generally in agreement with 
 those found in the large samples of \citet{zhu13} and \citet{rag16}.

\begin{figure}
\includegraphics[width=\columnwidth]{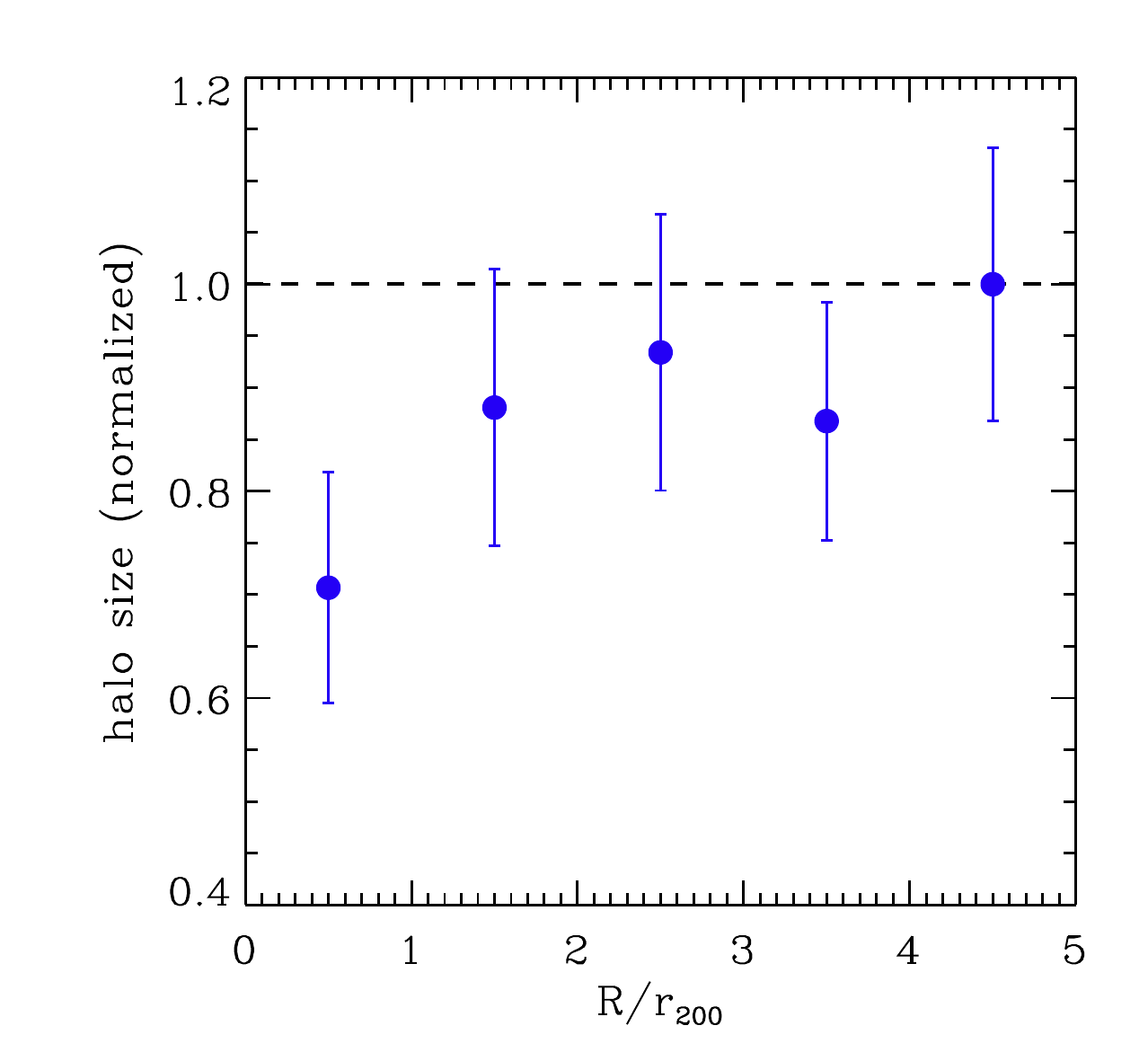}
\caption{Relative \MgII\ halo size variation on cluster-centric radius.
The horizontal dashed line is overplotted to guide the eye.
This result is based on the assumption that
\MgII\ absorbers are mostly associated with the CGM of bright-blue galaxies.
\label{fig-size}}
\end{figure}

\begin{figure*}
\includegraphics[width=0.85\linewidth]{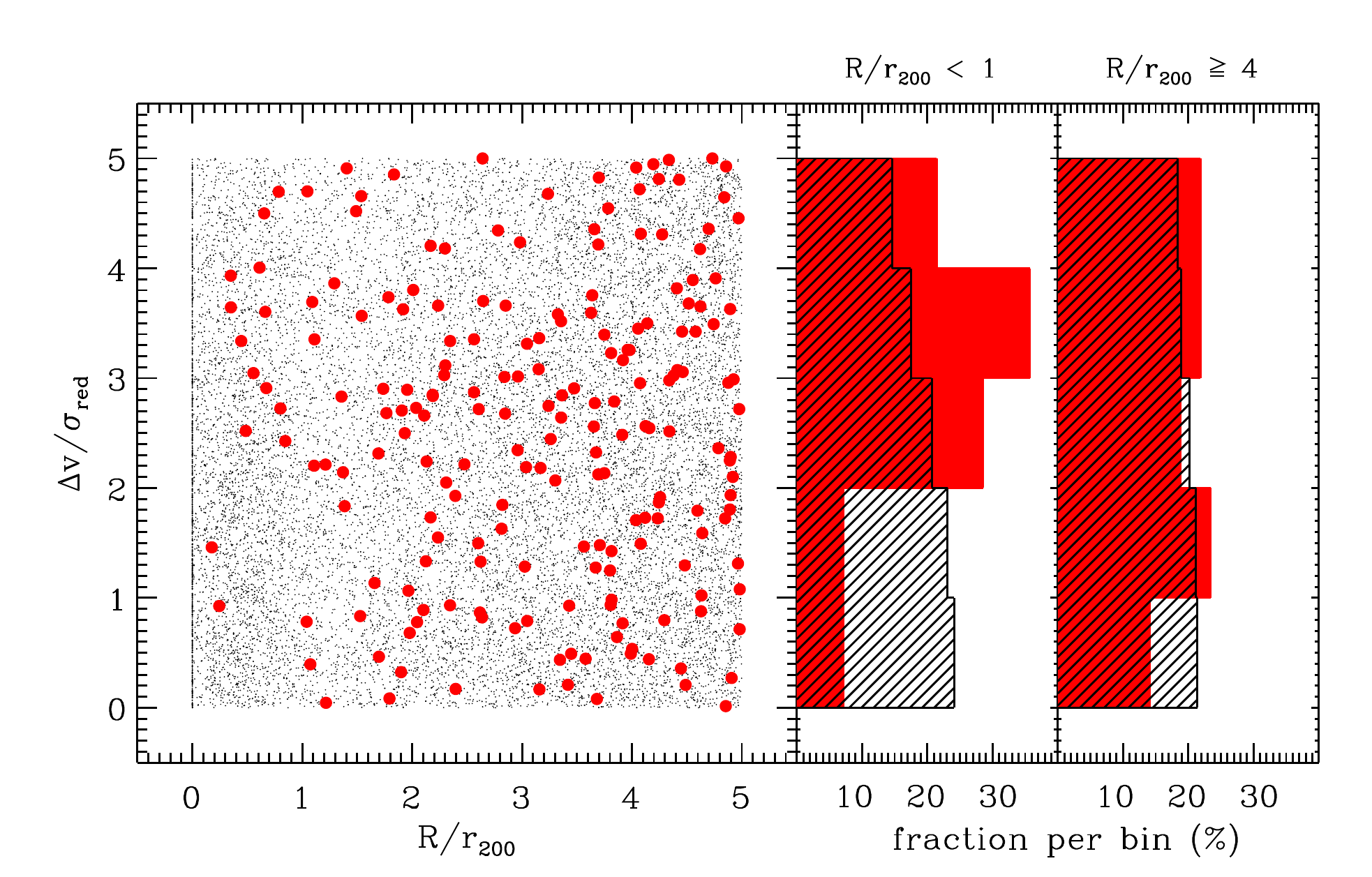}
\caption{Projected phase-space diagram of \MgII\ absorbers and
\redm\ cluster galaxies (left) and their $\Delta v$/\reds\ histograms
at $R/r_{200} <$ 1 (middle) and at $R/r_{200} >$ 4 (right). 
The black dots and hatched histogram represent the cluster galaxies,
while the red circles and shaded histogram are for the \MgII\ absorbers.
In the left panel, (randomly selected) 30 per cent of galaxies are presented 
for clarity.
\label{fig-phase}}
\end{figure*}

\subsection{Radial distribution in and around galaxy clusters} \label{radial}

In the left panel of Figure \ref{fig-rate},
 we present the \MgII\ detection rate per quasar,
 the number ratio of \MgII\ absorbers to quasars, in each $R/r_{200}$ bin.
Because a small number of quasars are matched with 
 distant clusters that are apparently small,
 the contribution from distant clusters is underestimated 
 when calculating the detection rate.
It is also underestimated in the clusters where the quasar number 
 density is low\footnote{The SDSS quasar survey is not uniform across the sky.
 The number density of quasars in the \redm\ cluster regions 
 varies from 18.8 to 89.1 deg$^{-2}$.}.
Therefore, we set the scaling factor for each cluster so that 
 the adjusted number of quasars matched with a cluster is 
 exactly proportional to the square of the cluster physical size
 and the total number of quasars remains the same.
The raw and corrected detection rates are 
 denoted by open and filled circles, respectively.
The reason why the corrected values are slightly high is because
 a large number of absorbers are detected at high-$z$ 
 due to the spectral characteristics mentioned above.
We also need to consider that the absorbers are easily detected 
 when using bright quasars.
However, the negligible dependence of quasar magnitudes on $R/r_{200}$, 
 as shown in Figure \ref{fig-distq}, enables us to assume that
 the radial trend of the detection rate reflects well
 the actual distribution of \MgII\ absorbers.
In this plot, we find that the detection rate increases 
 as the cluster-centric radius decreases: on average 
 0.59$\pm$0.13 per cent at $R/r_{200} <$ 1 and 
 0.22$\pm$0.03 per cent at $R/r_{200} \gid$ 4,
 indicating that \MgII\ absorbers are 2.70$\pm$0.66 times 
 more abundant in the clusters.
    
We show the normalized number density of cluster galaxies $\Sigma_{\rm gal}$
 in the right panel.
The bright-red, faint-red, bright-blue and faint-blue galaxies
 are denoted by the red, orange, blue and cyan squares, respectively.
The raw and spectroscopic completeness corrected values
 are presented in the open and filled symbols.
The radial variations of $\Sigma_{\rm gal}$ 
 depend on galaxy magnitude and colour:
 17.17$\pm$0.83, 7.14$\pm$0.41, 5.41$\pm$1.08 and 3.88$\pm$0.34
 times higher at $R/r_{200} <$ 1 than at $R/r_{200} \gid$ 4 for 
 the bright-red, faint-red, bright-blue and faint-blue galaxies,
 following the well-known relation that a large fraction of
 massive passive galaxies are in clusters \citep[e.g.][]{par09}.
Meanwhile, these values are larger than 2.70$\pm$0.66
 found in the \MgII\ absorbers.
If we assume that the majority of absorbers are connected to the CGM of 
 cluster galaxies in particular for the bright-blue galaxies,
 the absorber-to-galaxy density ratio at $R/r_{200} <$ 1 is  
 0.50$\pm$0.16 times that at $R/r_{200} \gid$ 4.
Because this ratio is proportional to the projected surface area of \MgII\ halo
 (i.e. ${\rm ratio} \propto {\rm halo~size}^2$),
 the value less than one means that 
 a typical halo size of cluster galaxies is smaller than that of field galaxies.
In Figure \ref{fig-size}, 
 we present the halo size normalized by that at $R/r_{200} \gid$ 4
 (the relative halo size hereafter)
 as a function of $R/r_{200}$ 
 based on the bright-blue galaxies.
The relative halo size is 0.71$\pm$0.11 at $R/r_{200} <$ 1.

This result is not as dramatic as the argument of \citet{pad09} that 
 \MgII\ halo radii can be changed from 50 to 10 kpc 
 by cluster environmental effects.
However, they might have overestimated the cluster effect
 in the sense that they did not take into account 
 the dependency of \MgII\ absorption on galaxy colour.
The relative halo sizes in the clusters become
 0.40$\pm$0.05, 0.62$\pm$0.08 and 0.83$\pm$0.11 
 when using the bright-red, faint-red, and faint-blue galaxies, respectively.
There is a possibility that we underestimate the cluster effect
 because weak absorbers (sensitive to environmental effects) 
 are not considered in this study,
 while \citet{pad09} utilized some weak absorbers from their own observations.

The truncated \MgII\ halo in cluster environment is supported 
 by the low covering fraction of Ly$\alpha$ absorption in the CGM of cluster galaxies 
 compared to field galaxies \citep[e.g.][]{yoo13,bur18}.
To explain this phenomenon, \citet{pad09} suggested the cold gas stripping scenario 
 by the hot intracluster gas. 
Because we cannot test this scenario with the current data sets,
 it is difficult to discuss the detailed mechanism of gas stripping:
 for example, is it driven by ram pressure \citep{gun72} or 
 thermal evaporation \citep{cow77}?
It is also difficult to rule out other types of processes
 such as strangulation \citep{lar80}, harassment \citep{moo96}, 
 tidal interaction with the cluster potential \citep{mer84} and
 high-speed multiple encounters with early-type galaxies \citep{par09}.

There are Ly$\alpha$ and \ion{C}{iv} absorbers 
 associated with the IGM component of clusters 
 rather than the CGM of cluster galaxies \citep[e.g.][]{yoo17,man19}.
However, \MgII\ absorbers can hardly trace the IGM
 because the \ion{H}{i} column density of IGM is too low,
 which is less than 10$^{16}$ cm$^{-2}$ in most areas 
 (see the hydrodynamic simulation result of \citealt{but19}).
The IGM contribution is expected to be small, 
 thus not changing our conclusion.

\subsection{Phase-space diagram} \label{phase}

Figure \ref{fig-phase} displays the distribution 
 of \MgII\ absorbers and \redm\ cluster galaxies in projected phase space.
The x-axis is the projected distance from the cluster centre 
 normalized by the cluster radius $R/r_{200}$, 
 while the y-axis is the line-of-sight velocity difference 
 between the absorber/galaxy and the cluster 
 normalized by the cluster velocity dispersion $\Delta v$/\reds.
The radial variation of spectroscopic completeness for the cluster galaxies 
 is not considered in the phase-space diagram.
For a quantitative comparison, we also present the $\Delta v$/\reds\ 
 histograms of the two populations at $R/r_{200} <$ 1 and $\gid$ 4.
The $\Delta v$/\reds\ distribution is 
 less affected by galaxy magnitude and colours, 
 so we include all the cluster galaxies in the histograms. 
Both histograms are nearly flat and identical 
 ($p$-value of the Kolmogorov--Smirnov test = 0.213) 
 in the field environment (i.e. $R/r_{200} \gid$ 4).
Inside the clusters, the cluster galaxies are naturally biased towards
 small $\Delta v$/\reds, even though the number of galaxies 
 with $\Delta v$/\reds\ $<$ 1 is only 1.66 times larger than that 
 with $\Delta v$/\reds\ $\gid$ 4:
 this may be due to non-negligible uncertainty 
 of \redz\ \citep{rin18,soh18}. 
Interestingly, the \MgII\ absorbers have rather large $\Delta v$/\reds.
As a result, the two histograms are significantly different 
 from each other ($p$-value = 0.010; 2.6$\sigma$ level) in the clusters.
It emphasizes that \MgII\ absorbers disfavour cluster environments.

If we recalculate the values obtained in Section \ref{radial} 
 using \MgII\ absorbers and cluster galaxies with $\Delta v$/\reds\ $<$ 1,
 the radial density variations of absorbers and bright-blue galaxies 
 become 1.33$\pm$1.18 and 5.88$\pm$2.64, respectively,
 leading to the change of relative halo size in the clusters 
 from 0.71$\pm$0.11 to 0.48$\pm$0.24.
This result shows that cluster environmental effects 
 can be evaluated more accurately within the narrower search window.
However, it is quite uncertain because there is only one absorber 
 at $R/r_{200} <$ 1 and with $\Delta v$/\reds\ $<$ 1.

\section{Summary} \label{summ}

Using the \redm\ galaxy clusters and SDSS DR14 quasars,
 we obtain 96 759 quasar-cluster pairs.
We then detect intervening \MgII\ absorption line features 
 in the background quasar spectra.
Our main results are as follows.
 
\begin{enumerate}
\item We find 197 \MgII\ absorbers with \ewa\ $>$ 0.3 \AA.
 The reliability of our identification is estimated to be $\sim$72 per cent.
 Their line parameters are not correlated with the cluster properties.
\item The \MgII\ absorber detection rate per quasar 
 is 2.70$\pm$0.66 times higher inside the clusters than outside the clusters.
 It shows that \MgII\ absorbers are abundant in clusters 
 compared to in fields.
\item Because the galaxy number density  
 is much higher inside the clusters than outside the clusters,
 the absorber-to-galaxy ratio is relatively low inside the clusters.
 We suggest that a typical halo size of (bright-blue) galaxies in clusters 
 is 0.71$\pm$0.11 times that in fields.  
\end{enumerate}
 
It is not easy to study environmental effects on the CGM in detail 
 with individual \MgII\ absorbers due to small number statistics.
Using stacked spectra of quasars around cluster galaxies,
 we plan to further investigate how the radial profile of
 \MgII\ absorption strength depends on galaxy and cluster properties.  
 
\section*{Acknowledgements}

We thank Daeseong Park for discussions in the continuum estimation.
We are grateful to an anonymous referee whose comments
helped to improve the original manuscript.

Funding for the Sloan Digital Sky Survey IV has been provided by
the Alfred P. Sloan Foundation, the U.S. Department of Energy Office of
Science and the Participating Institutions. SDSS-IV acknowledges
support and resources from the Center for High-Performance Computing at
the University of Utah. The SDSS web site is www.sdss.org.

SDSS-IV is managed by the Astrophysical Research Consortium for the 
Participating Institutions of the SDSS Collaboration including the 
Brazilian Participation Group, the Carnegie Institution for Science, 
Carnegie Mellon University, the Chilean Participation Group, 
the French Participation Group, Harvard-Smithsonian Center for Astrophysics, 
Instituto de Astrof\'isica de Canarias, The Johns Hopkins University, 
Kavli Institute for the Physics and Mathematics of the Universe 
(IPMU) / University of Tokyo, Lawrence Berkeley National Laboratory, 
Leibniz Institut f\"ur Astrophysik Potsdam (AIP),  
Max-Planck-Institut f\"ur Astronomie (MPIA Heidelberg), 
Max-Planck-Institut f\"ur Astrophysik (MPA Garching), 
Max-Planck-Institut f\"ur Extraterrestrische Physik (MPE), 
National Astronomical Observatories of China, New Mexico State University, 
New York University, University of Notre Dame, 
Observat\'ario Nacional / MCTI, The Ohio State University, 
Pennsylvania State University, Shanghai Astronomical Observatory, 
United Kingdom Participation Group,
Universidad Nacional Aut\'onoma de M\'exico, University of Arizona, 
University of Colorado Boulder, University of Oxford, 
University of Portsmouth, University of Utah, University of Virginia, 
University of Washington, University of Wisconsin, Vanderbilt University
and Yale University.

\section*{Data Availability}

The data underlying this article are available in the article and
in its online supplementary material.






\appendix

\section{Comparison with the JHU catalogue} \label{appen}

\begin{figure*}
\includegraphics[width=\linewidth]{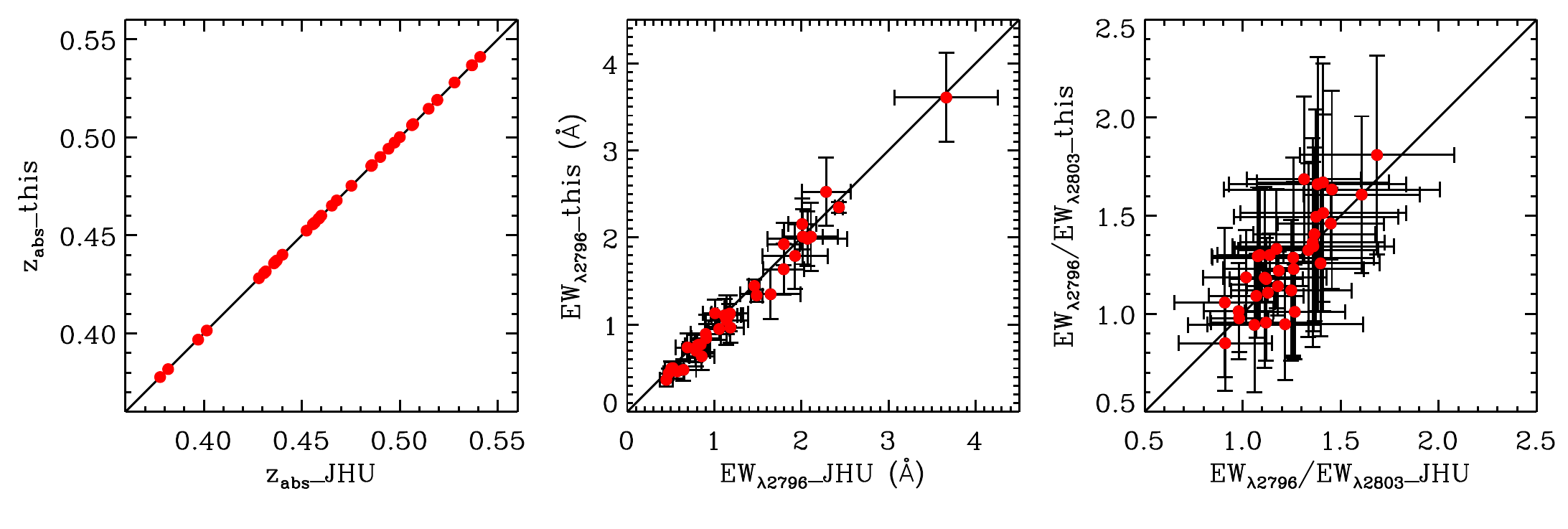}
\caption{Comparison of the measurements for \MgII\ absorbers
between this study and the JHU catalogue.
The left, middle and right panels are for
the redshift, EW of \MgII\ $\lambda$2796 and doublet line ratio, respectively.
The one-to-one relation (solid line) is overplotted.
\label{fig-a1}}
\end{figure*}

\begin{figure*}
\includegraphics[width=\linewidth]{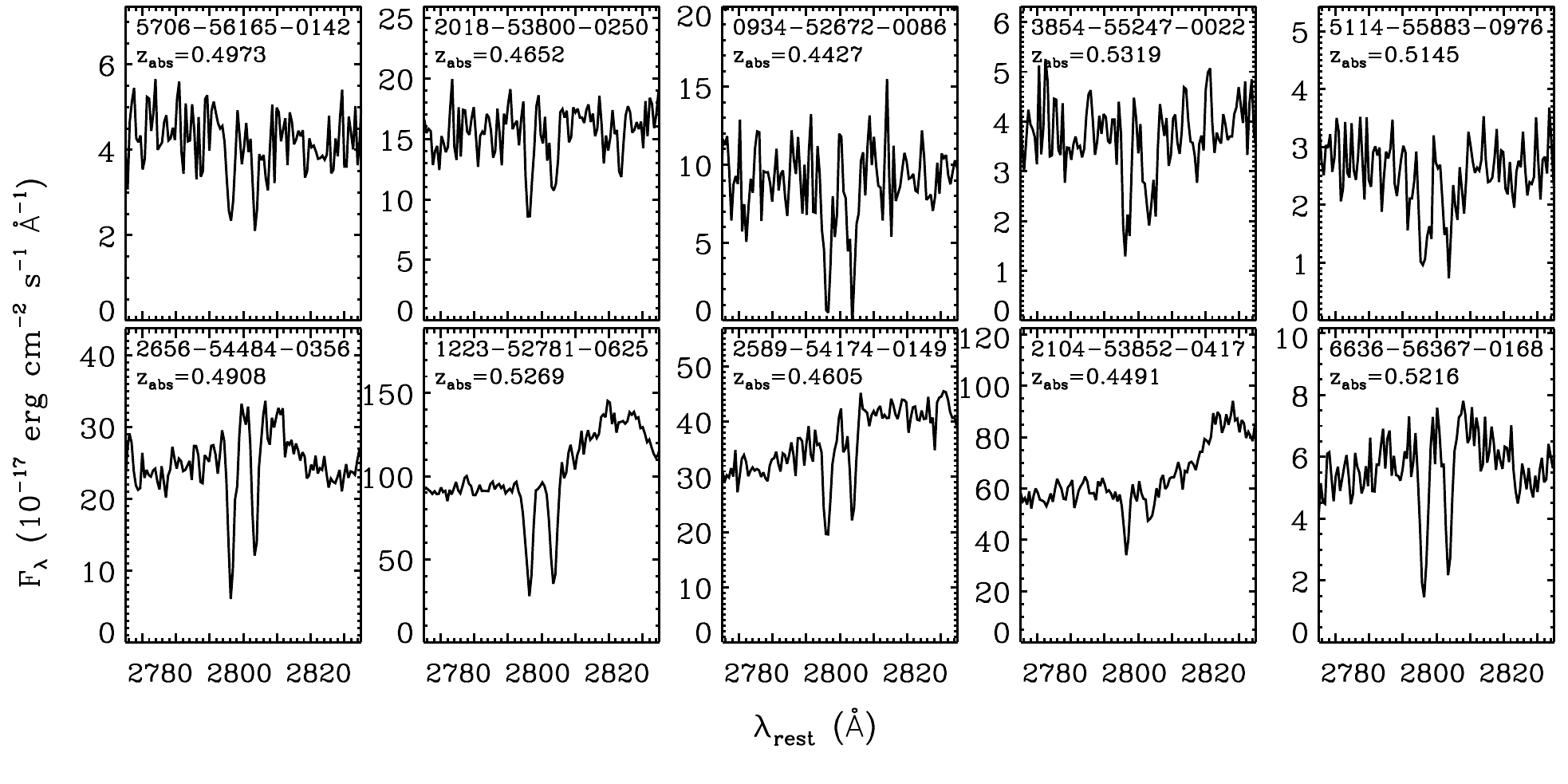}
\caption{Example spectra for the \MgII\ absorbers missed in this study (top).
The five cases with the highest S/N of total EW are presented.
Spectra of the five absorbers missed in the JHU catalogue (bottom).
The spectrum ID and absorber redshift are denoted in each panel.
\label{fig-a2}}
\end{figure*}

\citet{zhu13} provide a large sample of intervening \MgII\ absorbers.
Their data sets\footnote{
 https://www.guangtunbenzhu.com/jhu-sdss-metal-absorber-catalog}
 are updated using the SDSS DR7 and DR12,
 the JHU catalogue hereafter. 
In the JHU catalogue,
 we find 124 absorbers associated with the \redm\ clusters 
 (i.e. $R/r_{200} <$ 5 and $\Delta v$/\reds\ $<$ 5).
Among them, 67 absorbers satisfy the criteria of
 S/N $\gid$ 3 for both \ewa\ and \ewb\ in their calculation.
The 36 absorbers are also found in this study.
For these common absorbers,
 our measurements are fully consistent with
 those in the JHU catalogue, as shown in Figure \ref{fig-a1}.
The measurement errors are comparable with each other as well.
There are 31 absorbers not selected in this study.
These do not go through the automated pipeline mentioned in Section \ref{anal}
 mostly because the line strength is weak:
 the S/N of \ewa\ or \ewb\ is less than 3 in our calculation.
The example spectra of the absorbers missed in this study
 are presented in the top panels of Figure \ref{fig-a2}.
Although we choose the five cases with the highest S/N of total EW, 
 the spectra are quite noisy.

On the other hand, among the 197 absorbers in this study,
 80 absorbers are not included in the JHU catalogue
 due to the small sample of quasars used.
The 76 absorbers are outside the search window of \citet{zhu13}.
They searched for \MgII\ absorbers
 within a restrictive redshift range\footnote{
 Among the 197 absorbers in this study,
 only 64 absorbers are inside the search window of \citet{zhu13}.
 Their method is effective to minimize contamination
 by considering that $\Delta$separation/resolution distribution
 of the 64 absorbers is strongly concentrated at zero.
 However, it seems to eliminate a significant fraction of real absorbers.}
 to avoid false detections from 
 intervening \ion{C}{iv} and Milky Way \ion{Ca}{ii} absorption lines. 
Among the remaining 41 absorbers,
 36 absorbers are found in the JHU catalogue,
 but five ones are missed.
The spectra of the absorbers missed in the JHU catalogue 
 are shown in the bottom panels of Figure \ref{fig-a2}.
We check that these absorbers are not contaminated 
 by \ion{Fe}{ii} lines (see section 2.2.3 of \citealt{zhu13}).
It is beyond the scope of this study to understand
 why they could not find these absorbers (see also \citealt{zha19}).

\bsp	
\label{lastpage}
\end{document}